# 分簇 VLIW DSP 上特殊指令实现之乘法累加指令*

刘彬彬[1,2]， 郑启龙[1,2]

[1](中国科学技术大学 计算机科学与技术学院,安徽 合肥 230027)

[2](安徽省高性能计算重点实验室,安徽 合肥 230027)

通讯作者: 刘彬彬, E-mail: robbertl@mail.ustc.edu.cn

**摘 要**: BWDSP 是一款采用 VLIW 和 SIMD 架构,针对高性能计算领域而设计的 32 位静态标量数字信号处理器.针对其特殊的体系结构和特定的应用场景,设计了一系列相应的特殊指令.然而现有的编译框架,并没有对这些特殊指令提供支持.因此,在传统的 Open64 编译器的框架下,针对 BWDSP 结构特点,提出了一种实现特殊指令的算法,通过该算法实现了乘法累加指令,以提高具有乘法累加需求算法的性能.实验结果表明,针对相关算法,该实现能够在 BWDSP 上取得最高 8.85 倍的加速比.

## Realize special instructions on clustering VLIW DSP: multiplication-accumulation instruction

Binbin Liu[1,2], Qilong Zheng[1,2]

[1](School of Computer Science and Technology, University of Science and Technology of China, Hefei 230027 china)

[2](Key Laboratory of High Performance Computing Anhui Province, Hefei 230027 china)

**Abstract**: BWDSP is a 32bit static scalar digital signal processor with VLIW and SIMD features, which is designed for high-performance computing. Associated special instructions are designed for its special architecture and application scenarios. However, the existing compilation framework doesn't meet these special instructions. Therefore, in the context of traditional Open64 compiler, proposed a special instruction algorithm. Through this algorithm implements the multiplication-accumulation operation with BWDSP structure, to improve the performance of algorithms with multiply-accumulate requirements. Expermental results show that the algorithm, which can make an maximum of 8.85 speedup on BWDSP.

DSP 广泛应用在调制解调器、数字图片、多媒体计算等系统中.DSP 的应用和通用处理器有很大的不同,经常被应用在迭代和计算强度很大的数字信号处理中.DSP 最早引入实现快速乘法累加计算的硬件乘法器,该乘法器通常对应一条加速信号滤波的特殊指令—MACC 指令这条指令是根据滤波算法一个重要公式(1-1)而设计,应用 MACC 指令能大大加速滤波算法的运算速度.

$$x(n)=\sum_{i=0}^{n}x(i)\delta(n-i) \qquad (1\text{-}1)$$

为了提高特定应用环境下的运行速度,DSP 增加了许多特殊的指令和功能单元,这使得 DSP 的结构和通用处理器的结构有了很大的不同,也对我们实现这些特殊指令提出了很大的挑战.

本文对特殊指令—MACC 指令的实现进行讲解.根据目标 DSP 的多簇架构和 SIMD 编译优化,结合特殊指令的具体要求,在编译器后端,添加相关特殊指令,修改相应处理流程,以实现 MACC 指令.

本文第 1 节介绍目标 DSP—BWDSP 和相应编译器—Open64 编译器的基本框架.第 2 节提出了一种实现 DSP 上特殊指令的算法框架.第 3 节根据第 2 节的算法,结合 Open64 编译器和 BWDSP,介绍 MACC 指令的实现过程.第 4 节对 MACC 特殊指令的实现进行测试和结果分析.最后总结全文,并提出未来可以进一步提高优化的方面.

---

* 基金项目:核高基重大专项(2012ZX01034-001-001);

## 1 研究背景

BWDSP 是一款针对高性能计算领域而设计的 32 位静态标量 DSP,采用 VLIW(超长指令字)、SIMD(单指令多数据流)架构,其内核由 4 个构造完全一致的基本执行簇组成,分别为簇 x,簇 y,簇 z,簇 t, 结构如图 1 所示[2].每个簇有 8 个算术逻辑运算单元(ALU),8 个乘法器(MUL),4 个移位器(SHF),1 个特殊功能单元(SPU).此外,该处理器内部还有三个结构相同的地址产生器 u、v、w,用来执行地址运算指令和访存指令.

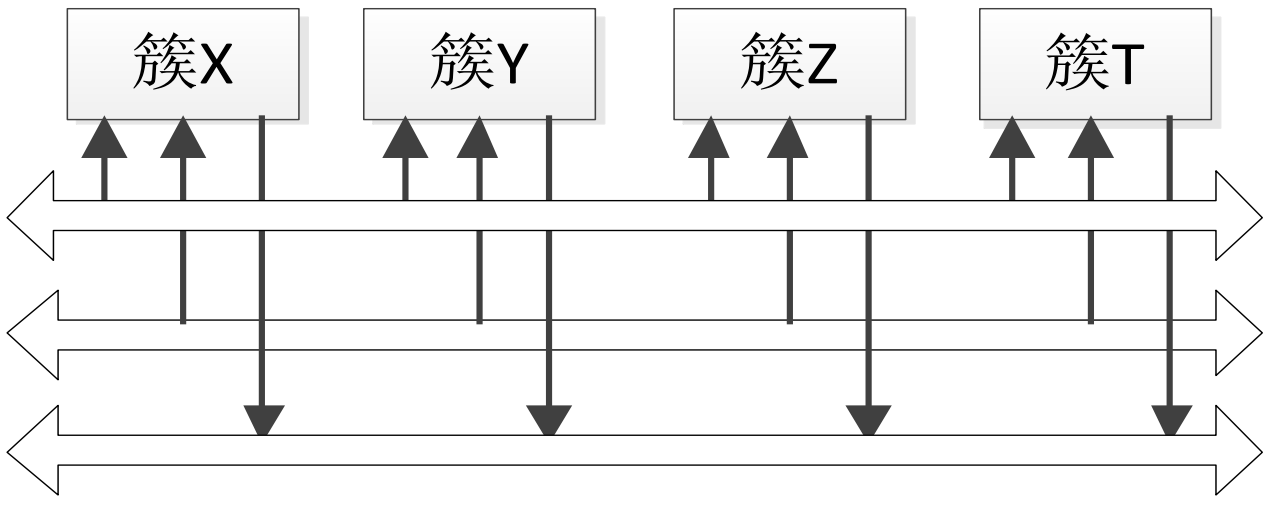

Fig.1 BWDSP structure

图 1 BWDSP 主要结构

BWDSP 编译器,是基于开源编译器 Open64[3]作为编译器研究框架,其主要工作在于后端重定向和针对特定体系结构的优化.Open64 完善的前端处理功能能够将源程序代码转化成相应的层次化的中间表示—WHIRL,支持全程序的分析优化,功能强大.同时又具有很好的移植性,后端支持多种体系结构,包括 MIPS、Itanium、ARM、AMD64、X86-64、IA32、NVIDIA 等.

Open64 的前端将源程序转化为中间表示 WHIRL[4],后端读入 WHIRL,翻译成代码生成阶段(Code Generation,CG)的中间表示 CGIR,在经过一系列优化,最终 CGIR 经过代码输出生成汇编程序[3].Open64 编译器的架构如图 2 所示.

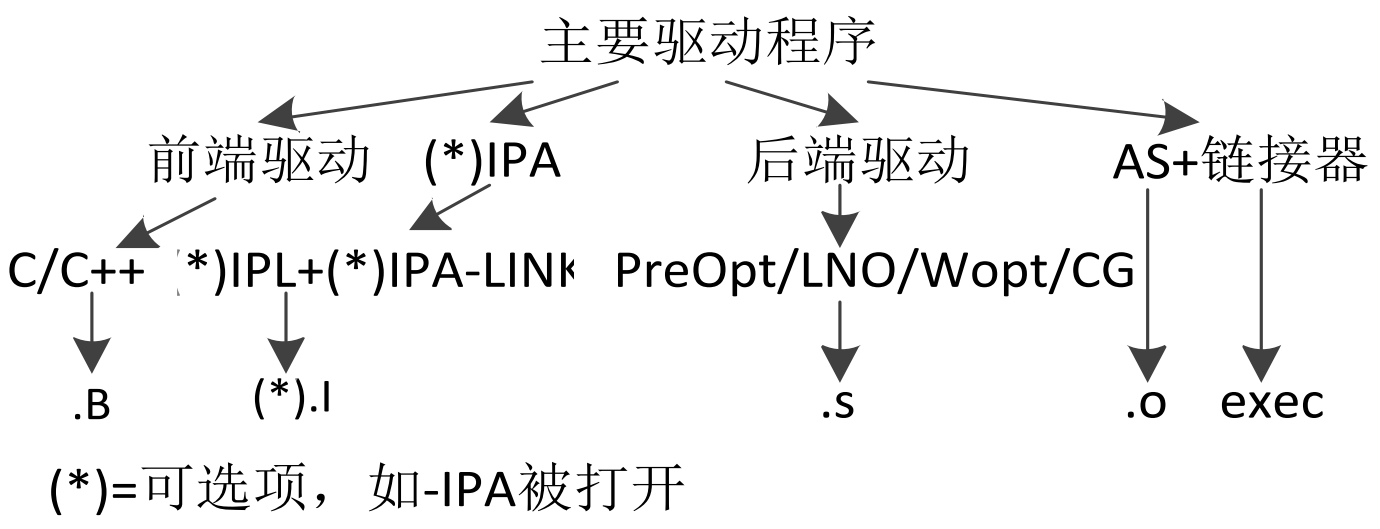

Fig.2 Open64 compilation model

图 2 Open64 编译架构

在 BWDSP 的指令集中,拥有 32 位定点乘法累加指令,双 16 位定点乘法累加指令,16 位定点复数乘法累加指令,32 位定点复数乘法累加指令[1].针对乘法累加指令(MACC 指令),BWDSP 每个簇拥有 4 个专用乘法累加寄存器(MACC0/1/2/3),用以加快乘法累加运算的速度.在 BWDSP 指令执行过程中,1 个乘法指令需要 3 个时钟周期,1 个加法指令需要 1 个时钟周期,1 个乘法累加指令仅需要 1 个时钟周期,故乘法累加指令的执行效率是非常高的.

## 2 特殊指令实现的算法框架

针对该体系结构的特性和特定的应用场景,设计了一系列特殊的指令,包括位反序寻址、乘法累加/累加指

令和移位器查找表指令等.结合 BWDSP 的特殊指令和 Open64 编译器架构,本文提出特殊指令的实现算法,包括以下 4 个步骤:

(1) 建立特殊指令的实现模型.根据特殊指令的具体情况,结合 Open64 编译器后端中间语言表示和处理流程,建立合适的特殊指令实现模型.

(2) 特殊指令的资源约束分析.特殊指令一般由专门的硬件驱动,由于资源的限制,特殊指令的实现也需要满足一定的约束条件.

(3) 在机器描述中加入特殊指令[11].Open64 编译器的机器描述采用二次编译形式,简化了对指令、cpu 建模的操作.对于特殊指令的实现,首先需要在机器描述文件中增加相应的指令描述,供编译器在运行时使用.

(4) 特殊指令的实现.在 Open64 编译器后端编译优化处理流程中,根据特殊指令的实现模型和相关的资源约束分析,实现特殊指令及其相应的处理流程

## 3 乘法累加指令的实现

由于 BWDSP 的特殊结构和乘法累加指令的特殊性,在对乘法累加指令实现的过程中,对编译器后端相关模块都有影响,因此需要进行相应的修改.相关模块包括机器描述、指令注释、寄存器分配和汇编指令生成等阶段.

本文选择在 Open64 编译器的中间语言 WHIRL 上,进行乘法累加指令的合成,参考本文提出的特殊指令实现算法,相应步骤如下.

### 3.1 建立乘法累加指令的实现模型

在编译器前端,1 条乘法累加指令,被拆分成 1 条乘法指令和 1 条加法指令.故乘法累加指令的查找模型为 opnd0=ADD(opnd0, MUL(opnd1, opnd2)),替换的结果为 MACC=opnd0, MACC +=opnd1*opnd2,opnd0=MACC.替换条件为:1.加法操作;2.一个操作数和一个乘法操作;3.输入的操作数是为了进行乘法操作;4.加法结果写回第一个操作数.

识别和替换模型如图 3 所示.

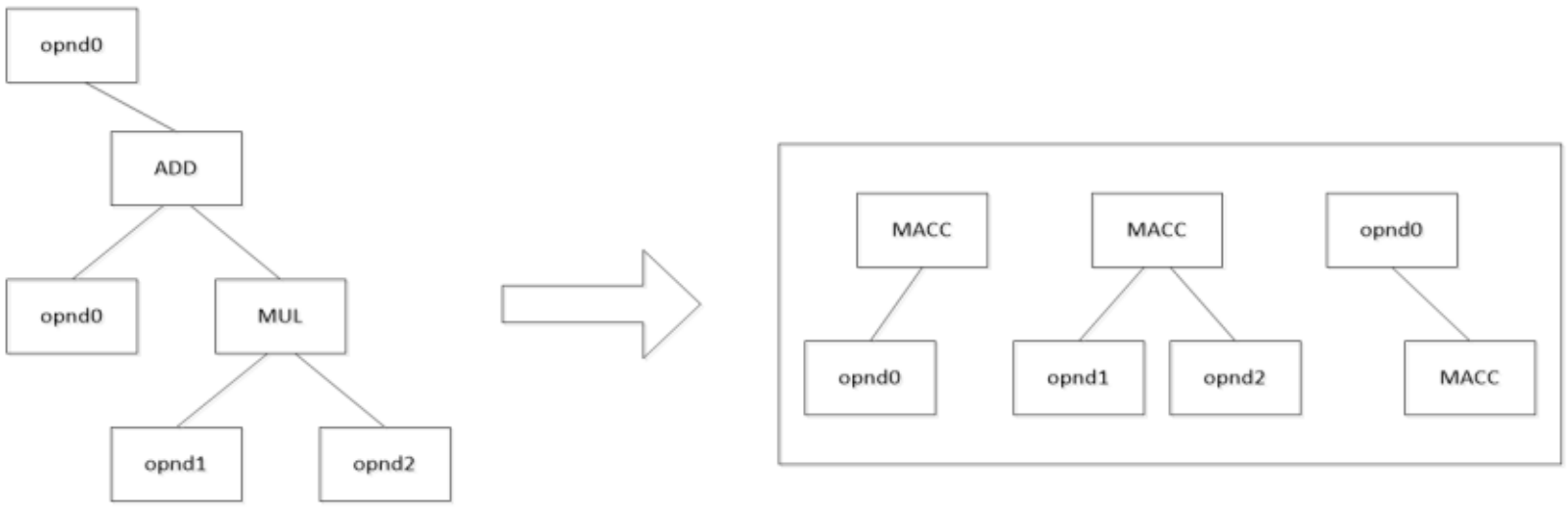

Fig.3 MACC instruction recognition and replacement algorithms

图 3 乘法累加指令识别和替换模型

### 3.2 乘法累加指令的资源约束分析

乘法累加指令是 SIMD 向量化[5]的一部分,SIMD 向量化位于编译器后端的循环嵌套优化中.由于 BWDSP 的每个簇上,只有 4 个乘法累加寄存器,故在对源程序的最内层循环进行 SIMD 向量化时,其乘法累加指令是不允许同时超过 4 条.如果超过了硬件资源的限制,则放弃合成.在满足硬件资源限制的条件下,选择优化性能更高、收益更大的双字指令模式进行 SIMD 向量化[6].

### 3.3 机器描述文件

Open64 编译器框架采用了二次编译的方式设计机器描述文件的架构,对乘法累加指令的支持通过填写相

应的机器描述[11]模型文件,完成 BWDSP 相应的寄存器模型后,在文件 bw104x-opcodes.knb 和 opcode_gen.cxx 中添加需要支持的相关乘法累加指令描述,然后编译机器描述文件,生成动态链接库供编译器在运行时使用.

修改汇编代码输出格式.对相关乘法累加指令按照 BWDSP 的指令格式进行汇编代码输出,汇编输出标志着编译器代码优化以及完成,是在 cgemit 阶段完成.在文件 isa_print.cxx 和 targ_info/isa/bw/isa_pack.cxx 文件中,增加相关指令的打印格式.

### 3.4 乘法累加指令的实现

#### 3.4.1 乘法累加指令的合成

编译器后端,将前端生成的 gspin tree 翻译为 high level whirl 树.在 whirl 树的 LNO(嵌套循环优化)处理阶段,自顶向下遍历最内层循环子树,根据乘法累加指令的识别模型,寻找满足条件的子树.

寻找到相关子树之后,进行相应的替换.在最内层循环前,将 MACC 寄存器赋初值为 opnd0.在循环中,合成 MACC 指令.在循环后,将 MACC 寄存器的值赋值给 opnd0.

#### 3.4.2 指令注释

中间语言 WHIRL 经过逐层下降之后,代码生成阶段调用指令注释将 WHIRL 树扩展成与机器指令一一对应的后端中间表示 CGIR 指令序列.由于指令注释依赖于目标处理器的机器指令,因此需要对 SIMD 指令注释进行特殊处理.具体的指令注释算法请参考文献[6].

指令注释示意图如图 4 所示.

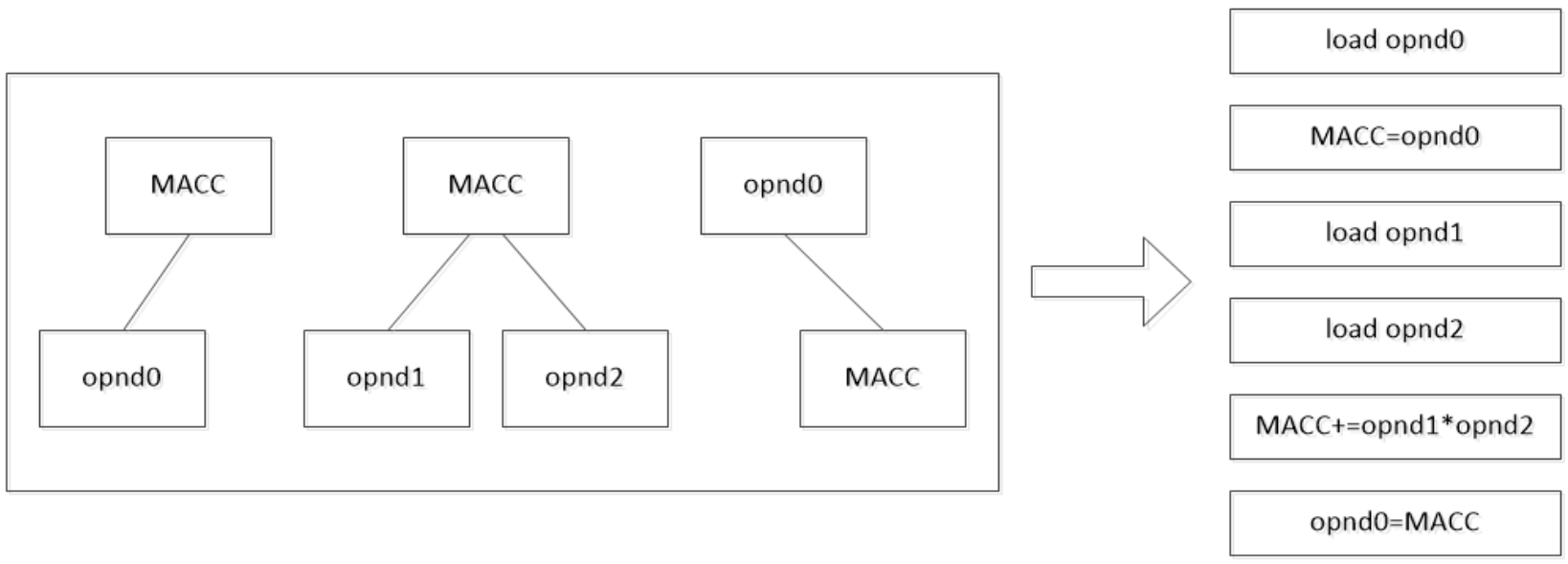

Fig.4 illustratiion of MACC instruction comments

图 4 乘法累加指令注释示意图

#### 3.4.3 指令分簇

指令分簇是指把每条 CGIR 指令指定到 x、y、z 和 t 相应簇上,通过多个簇同时执行来实现并行.具体的指令分簇算法请参考文献[7].

#### 3.4.4 寄存器分配

寄存器分配是为每个虚拟寄存器分配其运算簇上面的实际寄存器,包括全局寄存器分配和局部寄存器分配.

由于乘法累加指令在进行寄存器分配时,已经跨越基本块了,故在全局寄存器分配阶段对其进行寄存器的分配. Open64 全局寄存器分配采用图着色算法,同时加入了优先级考虑以及分裂活跃区间的概念[9].在进行全局寄存器分配的时候,优先考虑乘法累加等操作指令的处理,将 MACC 指令分配到专用的乘法累加寄存器中,然后再考虑其它指令的全局寄存器分配.局部寄存器分配采用更精确的线性分配方法,为基本块一级的局部变量分配寄存器[10].SIMD 指令寄存器分配主要在局部寄存器分配阶段,具体分配算法请参考文献[6].

#### 3.4.5 汇编指令生成

对相应的乘法累加指令,按照 BWDSP 的指令格式,进行汇编代码输出.

## 4 实验结果和分析

本文针对 BWDSP 的特定架构和特殊指令,实现了相关的乘法累加特殊指令.我们选取卷积运算、向量点积、fir 滤波算法和矩阵相乘,来验证方案的有效性.

在BWDSP编译器的开发平台环境Red Hat Enterprise Linux 5中对测试程序是否实现乘法累加指令进行测试,采用的性能指标为时钟周期数.运用 BWDSP 的调试器 ECS 来测量优化前后生成的汇编指令代码的时钟周期数.相关结果如表 1 所示.

**Table 1** Performance of MACC instruction

**表 1** MACC 指令性能

| 测试算法 | 实现前周期数 | 实现后周期数 | 加速比 |
|---|---|---|---|
| 向量点积(N=1024) | 17430 | 1969 | 8.85 |
| 卷积运算(N=1024) | 19508 | 3289 | 5.93 |
| fir 滤波(n=1024,N=128) | 2099277 | 428423 | 4.90 |
| 矩阵相乘(100*100) | 23101215 | 4176534 | 5.53 |

从上表可以看出,相关算法的加速比达到了 5.5~8.8 倍.下面以向量点积为例,进行更加详细的说明.

测试用例中的向量点积程序如下:

```
int sum=0
for(i = 0;i < N;i++){                //N=1024
   sum += a[i] * b[i];
}
```

未实现乘法累加指令前生成的汇编代码:

```
_Lt_0_3586:
xr11=[u5+=1,0]                    // u5 为数组 a 的地址
||xr13=1
||xr14=1023
xr10=[u6+=1,0]                    // u6 为数组 b 的地址
||xr15=r15+r13
xr11=r11*r10
.code_align 4
If xr15!=r14 B _Lt_0_3586
||   xr14=r14+ r11
_Lt_0_2050:
u6=__sum                          //u6 为 sum 的值
[u6+0,0]=xr14
```

实现乘法累加指令后生成的汇编代码:

```
_Lt_0_2562:
xyztr13=0
xyztMACC0=r13
xyztMACC1=r13
_Lt_0_770:
xyztr19:18=[u5+=8,1]              // u5 为数组 a 的地址
||xr17=8
||xr16=1023
```

```
xyztr21:20=[u6+=8,1]                    // u6 为数组 b 的地址
||xr15=r15+r17
xyztMACC0+=r19*r21
.code_align 4
If xr16>=r15 B _Lt_0_770
||xyztMACC1+=r18*r20
_Lt_0_1538:
u6=__sum
||xyztr8=MACC0
xyztr9=MACC1
xyztr8=r8+r9
xr10=sigma xyztr8
[u6+0,0]=xr10
```

从上面的汇编代码中可以看出,对于卷积运算,未实现乘法累加指令循环中的指令周期为 17*1024=17408,表中周期略高,因为包括其他一些指令周期在内.实现乘法累加指令后指令周期为 15*1024/8=1920,包括循环前的乘法累加寄存器赋初值和循环后乘法累加寄存器结果归约等指令周期,共为 1969.通过计算,其加速比为 8.85.

## 5 结语

本文针对 BWDSP 分簇体系结构及其特殊指令,在 Open64 编译器基础设施上,提出了一种特殊指令实现算法,并按照该算法,成功实现了其乘法累加特殊指令,并且在 BWDSP 编译器上对实现效果进行了测试和分析.实验结果表明,可取得最高 8.85 倍的加速比.今后,需要继续扩展软流水方法、零开销循环等优化技术,提升乘法累加优化应用性能.

**附中文参考文献:**